\documentclass[amsmath,amssymb,floats,preprint,showpacs,preprintnumbers]{revtex4}
\usepackage{graphicx}
\usepackage{dcolumn}
\usepackage{bm}

\def\be{\begin{equation}}
\def\ee{\end{equation}}
\def\bea{\begin{eqnarray}}
\def\eea{\end{eqnarray}}

\begin{document}
\title{Josephson effect for superconductors lacking center of inversion}
\author{Brigitte Leridon$^1$, Tai-Kai Ng$^2$ and C. M. Varma$^3$}
\affiliation{$^1$ UPR5 -  CNRS/ESPCI - 10 rue Vauquelin -75231 Paris cedex 05 - France \\ $^2$ Department of Physics -  HKUST - Hong Kong \\ $^3$ Department of Physics -  University of
California - Riverside -  CA 92521}
\begin{abstract}
Due to the absence of a center of inversion in some superconducting compounds, a p-wave admixture to the dominant d-wave order parameter must exist. If time-reversal is also violated, an allowed invariant is the product of the d-wave, p-wave and an appropriately directed current. We show that this leads to an anomalous Josephson current for tunneling along the direction parallel to the axis of the p-wave component, where the  current is the Meissner current and the Josephson loop current along the surface of the tunnel barrier. These ideas are applied to  the superconducting state of the cuprates in the pseudogap region of the phase diagram where in the normal phase some experiments  have detected a time-reversal and inversion symmetry broken phase.
 The effect is relevant also to heavy-fermion compounds which lack center of inversion due to crystalline symmetry.

\end{abstract}
\maketitle

A microscopic state violating time-reversal and inversion symmetry and reflection symmetry on some planes has been proposed for the pseudogap state of
the cuprates \cite{cmv}. The microscopic ground state associated with such symmetry breaking has spontaneous current-loops  in the O- Cu-O plaquettes in each cell  from zero temperature all the way to the pseudogap  temperature $T*$.
Two classes of experiments ( ARPES with circularly polarized photons \cite{kaminski, sv} and polarized neutron diffraction \cite{bourges}) on two
different cuprate compounds have observed such a state below $T*$. These broken symmetries are expected to continue into the superconducting state, as indeed
found in the experiments \cite{kaminski, bourges}. This should then also affect the symmetry of superconductivity \cite{ng, kaur}. Some heavy-fermion superconductors lack center of inversion in their crystal structure and similar superconducting states are expected for them \cite{bauer, agter}

Quite generally, if time-reversal
and inversion symmetry  are broken, a term in the free-energy of the form
 \be
i\epsilon(\psi_d^* \psi_{px} - c.c.) \label{dp-coup} \ee is
allowed.  $\psi_d$ is the dominant "d-wave" superconducting order
parameter, $\psi_{px}$ is a "p-wave" order parameter and the
 coefficient $\epsilon$ is proportional to the order parameter associated to the symmetry breaking in the
  normal phase (for example the crystal structure lacking center of inversion \cite{bauer} or the spontaneous current-loops \cite{cmv} in cuprates).  x corresponds to the $(110)$ direction or the $(1\bar{1}0)$ direction depending on the  domain.    We have introduced $i$ explicitly in the coefficient so that $\epsilon$ is real.
  In this case a superconducting state violating time-reversal and inversion symmetry and consistent with the proposed normal state is of the generic form
  $\psi(d_{xy})+i\delta_0\psi(p_{x,y})$
   in zero magnetic field, where the $(110)$ direction in the crystal has been taken as the $x$ direction. $\delta_0 = \epsilon/2\alpha_d$, where the leading term in the free-energy for the $d-$wave component is $\alpha |\psi_d|^2$.
 Such states have actually been investigated in context of some other materials whose crystal structure lacks inversion symmetry \cite{bauer, agter}.
Some experiments have already been proposed to discover evidence for such a superconducting state in the Cuprates \cite{ng}. Here we will show that the Josephson effect between such superconductors and a conventional $s-$wave superconductor with a particular orientation of the tunnel barrier with respect to the reflection symmetries of the superconducting state has a distinct signature from conventional Josephson effect. The clear observation of such effects will further substantiate the nature of the pseudogap phase for the cuprates and the nature of superconductivity in the relevant heavy-fermion compounds.

Consider a surface of the superconductor perpendicular to the x-direction $(110)$.
Then  for the state $\psi(d_{xy})+i\delta_0 \psi(p_{x})$, spontaneous super-current flows in the y-direction, i.e. along the surface, which is allowed. The alternative state $\psi(d_{xy})+i\delta_0 \psi(p_{y})$ has current normal to the surface and is not allowed.  So the presence of a surface whose normal is the x-direction will naturally favor the $p_x$ admixture.
Let us now consider that on this surface a  junction is fabricated which  allows a Josephson coupling  to a conventional s-wave superconductor.
We now wish to consider the modification of the superconducting
state due to an applied magnetic field in a direction in the plane of the junction. This modification must come about because beside the invariant of Eq.(1), there also exists an invariant proportional to $ij_y\psi(p_x)\psi(d_{xy})$. This is  because $j_y$, the current in the $y$-direction, is odd under time-inversion (as is $i\psi(p_x)$) and the product of the three terms satisfies all inversion and reflection invariances. Such a term can be derived from
the Ginzburg-Landau free-energy for the problem as follows:

Let the magnetic field ${\bf B}$ be in the $\hat{z}$-direction and adopt the London gauge in
which the vector potential is
\be
{\bf A}=B x \hat{y}.
\ee
The Free-energy density may then be written as
\bea
F(\psi(d_{xy}),\psi(p_x),A) = \alpha_d|\psi(d_{xy})|^2+K|\vec{D}\psi(d_{xy})|^2+\beta_d|\psi(d_{xy})|^4 +
\alpha_p |\psi(p_x)|^2 + \\
  i\epsilon(\psi(d_{xy})^*\psi(p_x)-\psi(d_{xy})\psi(p_x)^*)+i\epsilon'(\psi(d_{xy})^*D_y\psi(p_x)-\psi(d_{xy})D_y^*\psi(p_x)^*),
\eea where $\vec{D}=-i(\nabla-2ei\vec{A}/\hbar c)$. This is simply
the modification of the free-energy written down by Agterberg and
Kaur \cite{kaur} for the present geometry and in the presence of a
magnetic field. $\epsilon$ has already been discussed; $\epsilon'$
has a magnitude of the order of the fermi-velocity and is
independent of the underlying inversion/time-reversal breaking
order parameter. On minimizing with respect to both the magnitudes
and phases of $d_{xy}$ and $p_x$, one obtains that the order
parameter is
 \be 
 \psi = \psi(d_{xy}) +i\delta \psi(p_x), 
 \ee where
\be 
\psi(d_{xy}) = |\psi(d_{xy})|\exp{(i\theta_{0x}+i\phi_B(x))}
,~
\phi_B(x)=\frac{2e}{\hbar}y\int^xB(x')dx' 
\ee 
and
 \be 
 \delta =[\delta_0 - \zeta j_y] \Delta_d\exp{i\phi_B(x)}. 
\ee 
Here
$\Delta_d$ is the magnitude of the major order parameter, $\theta_{0x} =x\epsilon\epsilon'/K^R$ specifies the spiral state deduced by Agterberg and Kaur \cite{kaur}, $\delta_0 =
(\epsilon(1+\epsilon'^2/K^R))$ is magnitude of the bulk induced $p-$wave order parameter and the  coefficient $K^R=K-\epsilon'^2/\alpha_p$ and $\zeta =
\epsilon'/(2\alpha_p\Delta_d^2K^R)$. The length associated with
the pitch of the spiral can be estimated to be much larger than
the London penetration depth $\lambda$. The spiral then produces
negligible effects in Josephson tunneling and will be ignored by
putting $\theta_0=0$.

The second part of Eq.(7) is equivalent to having the aforementioned term in the free-energy  proportional to $ij_y\psi(p_x)\psi(d_{xy})$. There are two contributions to $j_y$. One is just the Meissner screening current, which is uniform on the surface,
\be
j_y^{(1)}=\frac{-cB}{4\pi\lambda}\exp{(-x/\lambda)}.
\ee
The other arises when a Josephson junction is constructed of this superconductor with another superconductor (of, for example the s-wave kind) and a magnetic field is applied in the junction as above. Then, the Josephson current density across the tunnel junction, $J_x(y)$ is periodic in $y$,
\be
J_x(y) = J_{x0} \sin(B w y/\Phi_0 +\gamma_0).
\ee
Here $w$ is the effective thickness of the barrier in the $x-$direction and $\gamma_0$ is the phase difference across the barrier. It then follows, by continuity or by considering the gradient of the phase on the surface of either superconductor, that there exists also a periodic current
\be
j_y^{(2)}(y) = J_x(y).
\ee
Such a periodic surface current is of-course present in every Josephson junction. Usually, it has no consequence. In the case of the $d+ip$ superconductor, it changes the magnitude of the $p_x-$wave component periodically in the $y-$direction.
Note that since this modulation is itself proportional to the Josephson current, it will have an effect on the Josephson current which is second order in the tunneling. We mention it for completeness although it has a negligible effect on our results and is dropped below.

Collecting the  results above, the wave-function of the $d+ip$ superconductor near the surface of the Josephson junction up to a depth about $\lambda$ in x-direction  is
\bea
\psi(y) & = &\Delta_d(f_d(\theta) + i \delta_y \cos(\theta)) \exp{(i\phi_B(y))}  \\  \nonumber
           &= & |\Delta(y)|\exp{i(\phi_B(y) +\phi(y))},
 \eea
 where
 \bea
  \Delta=\Delta_d\sqrt{f_d^2(\theta)+\delta^2(y)\cos^2(\theta)} \\
   \phi = \arctan (\delta \cos \theta/f_d(\theta)) \\
   \delta = (\delta_0 + \delta_1 B), ~ \delta_1=\zeta c/(4\pi \lambda).
   \eea

We may now calculate the Josephson current between the $d+ip$ superconductor oriented along the $x$ or $(110)$ direction and a conventional s-wave superconductor.
    \be
I_J  \propto \int_{-\pi/2}^{\pi/2} d\theta F(\theta)\int_{-L/2}^{L/2} dy \Delta
 \sin [(\gamma_0 + \frac{2e}{\hbar} B\lambda y + \phi(y)].
 \label{J-current}
 \ee
 $F(\theta)$ gives the tunneling cone
 and $L$ is the length of the barrier in the $y-$direction, i.e. along $(1\overline{1}0)$ .  We can evaluate the integral to get
\be
I_J=A_1\int^{L/2}_{-L/2}dy\left(\sin(\gamma_0+(2\pi{B}\lambda_L/\Phi_0)y)\right)
+A_2\int_{-L/2}^{L/2}dy\left(\delta\cos(\gamma_0+(2\pi{B}\lambda_L/\Phi_0)y)\right),
\ee
 where $A_1=\Delta_0\int_{-\pi/2}^{\pi/2}d\theta F(\theta)f_d(\theta)$ and $A_2=\Delta_0\int^{\pi/2}_{-\pi/2}d\theta
F(\theta)\cos(\theta)$. We obtain
 \be
 {I_J\over L}=A_1\sin(\gamma_0){\Phi_0\over\pi\Phi}\sin({\pi\Phi\over\Phi_0})+A_2\left((\delta_0+\delta_1B)\cos(\gamma_0)
{\Phi_0\over\pi\Phi}\sin({\pi\Phi\over\Phi_0})\right)
 \label{j-current2}
 \ee
 $\Phi=B\lambda_LL$ is the total flux through the junction.
 This expression should be maximized with respect to $\gamma_0$ to get the observable Josephson current.
 The $I_J$ thus obtained is
 \be
 {I_J\over L}=\left[\left(A_1{\Phi_0\over\pi\Phi}\sin({\pi\Phi\over\Phi_0})\right)^2+\left(A_2(\delta_0+\delta_1B){\Phi_0\over\pi\Phi}
 \sin({\pi\Phi\over\Phi_0})\right)^2\right]^{1\over2}
 \label{j-current3}
 \ee
  The first term in (\ref{j-current3}) is the usual term; It would be zero in a tetragonal crystal where $f_d(\theta)=\sin(2\theta)$.  The second term has a part due to the p-wave admixture; the more interesting part of it is proportional to B arising from the Meissner current, which cancels the flux $\Phi$ in the denominator to give a part which as a function of B is oscillating with a constant amplitude. This last feature is remarkable since it will lead to a non decreasing amplitude for the Josephson currents with magnetic field for large fields. This happens due to quite different physics in squids and at a quite different period.

  The results of the evaluation of Eq. (18) are represented on figure 1, for a conventional $d$ or $s$ or $d+s$ material  (red dotted line), and  a $d+ip$ material  (blue solid line). The main observation is that the critical current does not decrease at large fields but oscillates.  For this calculation it has been assumed that a single domain of the $p-$wave component exists over the entire length L of the junction.

 \begin{figure}
\begin{center}
\includegraphics{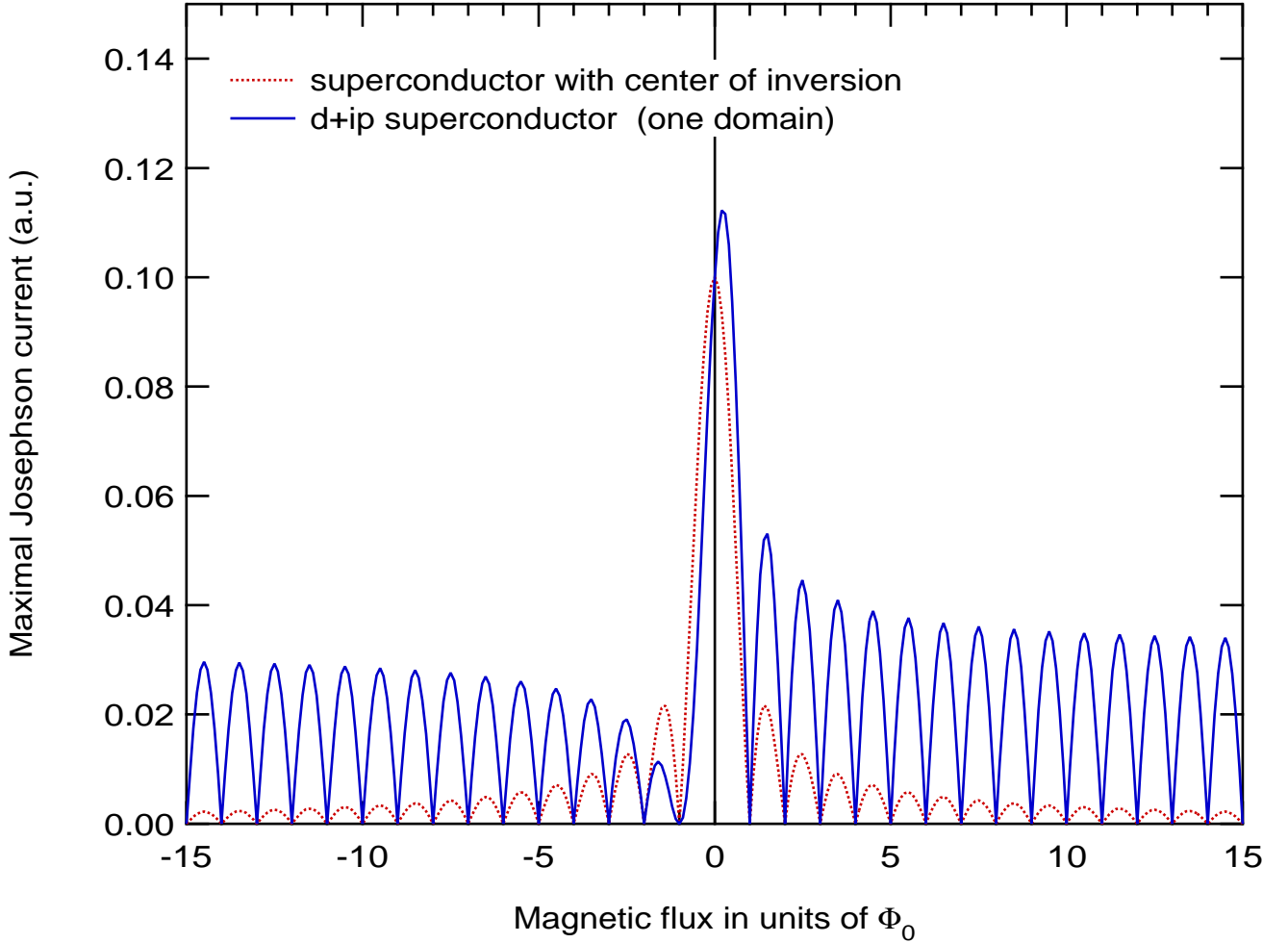}
\caption{(Color on line.)  Josephson current patterns as a function of magnetic field. Red dotted line: Conventional Josephson pattern expected for a superconductor.  Blue solid line: Josephson pattern expected for a single domain $d+ip$ superconductor.  The parameters used for the calculation in reference to Eq.(17) are $A_1 = 0.02 A.m^{-1}, A_2\delta_0 = 0.1 A.m^{-1},
A_2\delta_1 = 0.05 A.m^{-1}.T^{-1}$. }
\label{default}
\end{center}
\end{figure}

  However, in real crystals,  d+ip domains are expected to occur, whenever the $p_x$ component changes sign with respect to the $d$ component, which introduces two other terms that were previously cancelled when integrating over the junction.
  Let us consider $n_{max}$ randomly distributed domains along the y-direction.  Each domain extends from $\frac{l_{n}}{L}$ to $\frac{l_{n+1}}{L}$. Each domain boundary  corresponds to a change of sign of the $p_x$ component.  In the  expression for the current, Eq. (18), the term proportional to $A_2$ is replaced by

$   -A_2\sum_{n=0}^{n_{max}}(-1)^{n+1}\times  $ \ $   $\\
 $     $  \qquad  $ (\delta_0+\delta_1B)\left[{\Phi_0\over2\pi\Phi}\left(\cos({2\pi\Phi\ell_{n+1}\over\Phi_0 L})-\cos({2\pi\Phi \ell_{n}\over\Phi_0 L})\right) + {\Phi_0\over\pi\Phi}
 \left(\sin({\pi\Phi\ell_{n+1}\over\Phi_0 L})-\sin({\pi\Phi\ell_{n}\over\Phi_0 L})\right) \\
 \right] \\ $

  The results of the calculation for multidomains samples are presented on figure 2 for two domains in two different configurations (red dotted line and blue dashed line) and 5 domains (black solid line). The usual Fraunhofer pattern is replaced by a complicated interference pattern. The patterns are not even in field and they do not show a  decrease with magnetic flux, however the periodicity of a flux quanta is still apparent.  (The modification of the pattern due to neglected Josephson loop currents introduces a second harmonic but their magnitude is expected to be very small.)   The observed experimental pattern should vary depending on the geometry of the domains, but in every case it should retain a component whose magnitude does not decrease with field at large fields.

 \begin{figure}
\begin{center}
\includegraphics{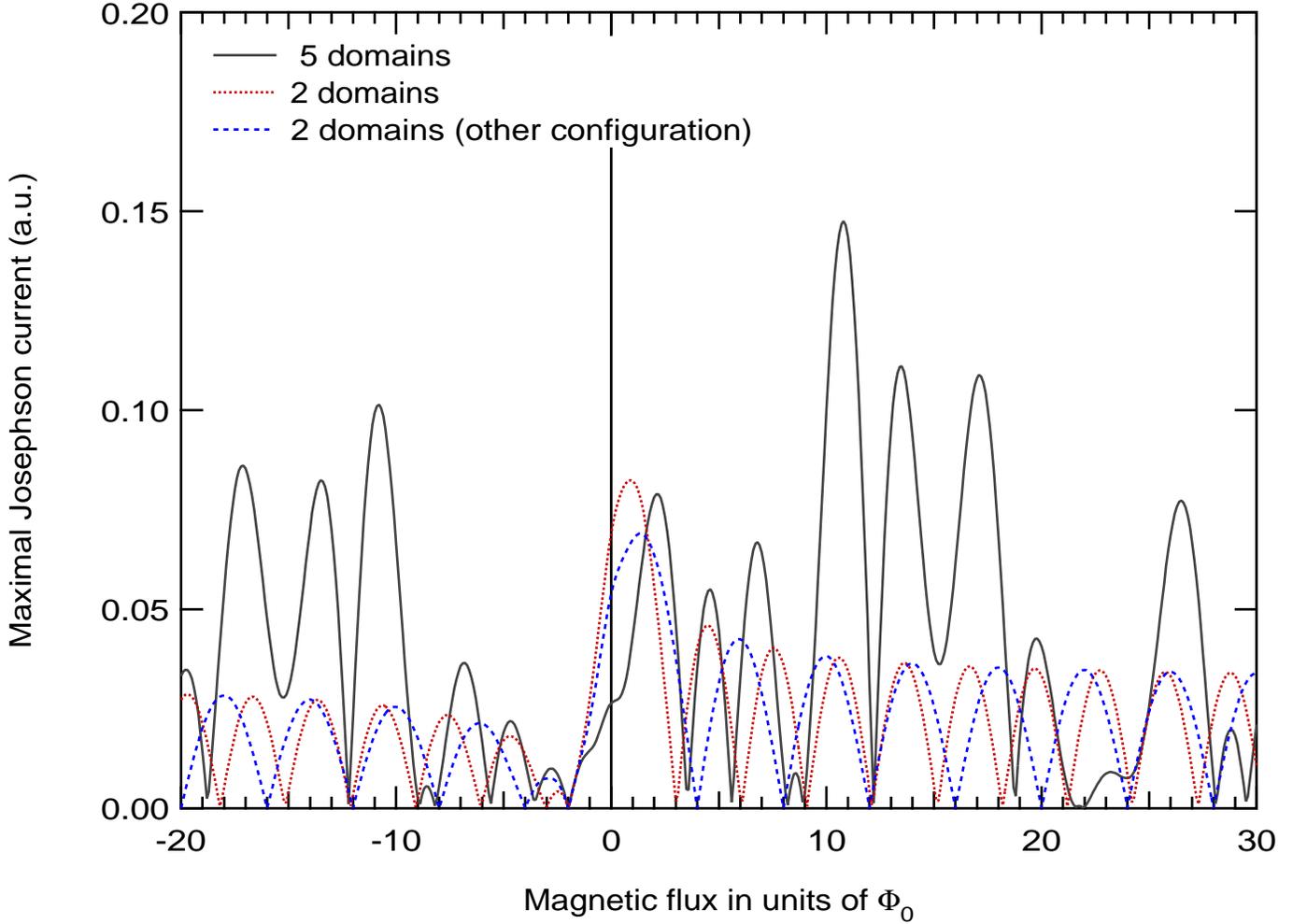}
\caption{(Color on line.)  Josephson current patterns as a function of magnetic field for a "$d+ip$" superconductor with domains of the $p-$wave component. The parameters are the same as in Fig.(1). Red dotted and blue dashed curves are for two domains of different configuration whose borders are given by (-L/2,-L/3,+L/2) and (-L/2,+L/4,+L/2) respectively. The black solid curve is for five random domains (-L/2, -0.41L, 0.09L, 0.15L, 0.23L, +L/2). Parameters are $A_1 = 0.02 A.m^{-1}, A_2\delta_0 = 0.1 A.m^{-1},
A_2\delta_1 = 0.05 A.m^{-1}.T^{-1}$. }
 \label{default}
\end{center}
\end{figure}

  When tunneling along the (001) direction of the cuprates, i.e. tunneling perpendicular to the Cu-O planes, the $p-$order parameter with or without a magnetic field contributes no Josephson current. Therefore in this geometry only a conventional Fraunhoffer pattern should be observed.

 The situation is more complicated when tunneling along the 100 direction, i.e. along the lobes of the $d-$wave order parameter. A $p-$wave component would then be necessarily accompanied by a current at an angle $\pi/4$ to the boundary. This is not allowed. A systems of domains along the surface can be envisaged which cancel the component of the current perpendicular to the surface. We have not investigated the energetics of this possibility, but if this is realized, the Josephson current variation with a magnetic field is expected also not to be decaying as in a Fraunhoffer pattern, due to the boost of the $p-$wave by the Meissner current.


 We are aware of one set of Josephson experiments in $YBa_2Cu_3O_{6+x}$ with tunneling direction perpendicular to the layers as well as along the $110$ and $100$-direction \cite{grison}. In the former the usual Fraunhofer diffraction pattern is indeed observed. In the latter, an oscillatory form remains at large $\Phi/\Phi_0$, which is consistent with the prediction here and qualitatively resembles the simulation on figure 2.  The exact pattern is expected to depend on the precise geometry of the $d+ip$ domains and therefore is expected to vary only upon warming up and cooling down the sample which was indeed observed.  However, in these experiments, it is not known (although probable) whether the thin films were underdoped. We suggest experiments in which the quasi-particle tunneling above Tc should be monitored simultaneously with the Josephson tunneling below Tc to see if the effects predicted here only occur in samples which show the pseudogap in the quasi-particle tunneling.

This experiment if verified may constitute an additional observation of the symmetry breaking in the pseudogap phase of the cuprates.  The predictions here also have applications in appropriate geometries to other superconductors lacking a center of inversion \cite{bauer}.

\begin{acknowledgments}
T.K. Ng acknowledges support by HKRGC through
grant number 602803.
\end{acknowledgments}

   \end{document}